\documentclass[aps,amsmath,amssymb,prl,twocolumn,superscriptaddress,letterpaper,10pt]{revtex4-1}
\usepackage{graphicx,color}
\usepackage{bm}
\usepackage[bbgreekl]{mathbbol}
\usepackage{txfonts}
\usepackage{mathrsfs}
\usepackage{feynmf}
\usepackage{comment}
\usepackage{epsfig}
\usepackage{setspace}
\newcommand{\al}{\alpha}
\newcommand{\be}{\beta}
\newcommand{\g}{\gamma}
\newcommand{\de}{\delta}

\newcommand{\thi}{\theta}

\newcommand{\p}{\pi}

\newcommand{\s}{\sigma}

\newcommand{\W}{\Omega}
\newcommand{\De}{\Delta}

\newcommand{\pd}{\partial}

\newcommand{\round}[1]{\left( #1 \right)}

\newcommand{\cvec}[2]{\round{\begin{array}{c} #1 \\ #2 \end{array}}}

\newcommand{\mat}[4]{\left(\begin{array}{cc}#1&#2\\#3&#4\end{array}\right)}

\newcommand{\beq}{\begin{equation}}
\newcommand{\eeq}{\end{equation}}
\newcommand{\Beq}{\begin{eqnarray}}
\newcommand{\Eeq}{\end{eqnarray}}
\newcommand{\bml}{\begin{multline}}

\newcommand{\bea}{\begin{align}}
\newcommand{\ena}{\end{align}}
\newcommand{\bsp}{\begin{split}}
\newcommand{\esp}{\end{split}}
\newcommand{\down}{\downarrow}
\newcommand{\up}{\uparrow}

\newcommand{\bS}{{\boldsymbol{S}}}

\newcommand{\ez}{{\boldsymbol e_z}}

\renewcommand{\bm}{{\boldsymbol m}}

\newcommand{\bn}{{\boldsymbol n}}

\newcommand{\hS}{\hat{S}}

\newcommand{\bx}{\boldsymbol{x}}

\newcommand{\ve}{\varepsilon}

\newcommand{\bB}{\boldsymbol{B}}

\begin{document}
\title{Spin Superfluidity in the $\nu=0$ Quantum Hall State of Graphene}
\author{So Takei}
\affiliation{Department of Physics and Astronomy, University of California, Los Angeles, CA 90095, USA}
\author{Amir Yacoby}
\affiliation{Department of Physics, Harvard University, Cambridge, MA 02138, USA}
\author{Bertrand I. Halperin}
\affiliation{Department of Physics, Harvard University, Cambridge, MA 02138, USA}
\author{Yaroslav Tserkovnyak}
\affiliation{Department of Physics and Astronomy, University of California, Los Angeles, CA 90095, USA}
\date{\today}

\begin{abstract}
A proposal to detect the purported canted antiferromagnet order for the $\nu=0$ quantum Hall state of graphene based on a two-terminal spin transport setup is theoretically discussed. In the presence of a magnetic field normal to the graphene plane, a dynamic and inhomogeneous texture of the N\'eel vector lying within the plane should mediate (nearly dissipationless) superfluid transport of spin angular momentum polarized along the $z$ axis, which could serve as a strong support for the canted antiferromagnet scenario. Spin injection and detection can be achieved by coupling two spin-polarized edge channels of the $|\nu|=2$ quantum Hall state on two opposite ends of the $\nu=0$ region. A simple kinetic theory and Onsager reciprocity are invoked to model the spin injection and detection processes, and the transport of spin through the antiferromagnet is accounted for using the Landau-Lifshitz-Gilbert phenomenology.
\end{abstract}
\maketitle

\singlespacing

{\em Introduction}.|Unique electronic properties of graphene (a monolayer of graphitic carbon) stem from its hexagonal lattice structure, engendering relativistic effects at electronic velocities well below the speed of light~\cite{castroRMP09,*sarmaRMP11}. Graphene is the thinnest and the strongest of 2D materials, and an outstanding electrical and heat conductor, holding great promise as a building block for future electronic devices~\cite{schwierzNATN10,*editorialNATN14}. 

A hallmark of graphene's electronic properties is manifested in magnetotransport. The integer quantum Hall (QH) sequence with anomalous filling fractions $\nu=\pm4(n+1/2)$~\cite{novoselovNAT05,*zhangNAT05,*castroRMP09,*sarmaRMP11} directly reflects the weakly-interacting massless relativistic (``Dirac") nature of its low-energy excitations and the fourfold degeneracy associated with the electron spin and valley isospin. The valley degree of freedom distinguishes between the two inequivalent ``Dirac points" in the Brillouin zone where the conduction and valence bands of graphene touch~\cite{zhengPRB02,*gusyninPRL05,*peresPRB06}. 

Under high magnetic fields, strongly-correlated QH phases can emerge~\cite{zhangPRL06,*checkelskyPRL08,*duNAT09,*bolotinNAT09,*zhangPRB09,*deanNATP11,*ghahariPRL11,*youngNATP12,youngNAT14}, including the $\nu=0$ state at the charge neutrality point, which indicates that interaction-induced SU(4)-symmetry breaking within the spin-valley space lifts the fourfold degeneracy of the zeroth Landau level~\cite{yangPRB06,*goerbigPRB06,*gusyninPRB06,*nomuraPRL06,*aliceaPRB06,*abaninPRL06,*fuchsPRL07,*shengPRL07,*abaninPRL07,*jungPRB09,*nomuraPRL09,*houPRB10,*abaninPRB13,herbutPRB07,*kharitonovPRB12,*kharitonov2PRB12}. A challenge is to understand precisely how this symmetry is broken for the $\nu=0$ state. Charge-transport experiments, utilizing both the two-terminal and Hall-bar geometries, suggest that the bulk and edge charge excitations for the state are gapped~\cite{zhangPRL06,*checkelskyPRL08,*duNAT09,*bolotinNAT09,*zhangPRB09,*deanNATP11,*ghahariPRL11,*youngNATP12}. Furthermore, the recent observation of gapless edge-state reconstruction in tilted magnetic field~\cite{youngNAT14} is consistent with the scenario where the $\nu=0$ ground state is a canted antiferromagnetic (CAF) insulator~\cite{herbutPRB07}, where the spins $\bS_A$ on sublattice $A$ have a different orientation relative to the spins $\bS_B$ on sublattice $B$; in the presence of an external magnetic field $\bB$ normal to the graphene plane (defined to be the $xy$ plane), the total spin $\bS_A+\bS_B$ points antiparallel to the field while the N\'eel vector $\bS_A-\bS_B$ lies in the graphene plane. Despite these recent developments, a more direct experimental verification of this CAF scenario would be highly desirable. 

Essentially disjoint from the field of graphene QH physics, spintronics is witnessing an increasing interest in realizing spin transport through magnetic insulators via coherent collective magnetic excitations, which allow for superfluid (nearly dissipationless) transport of spin~\cite{konigPRL01,*soninAP10,*takeiPRL14,*chenPRB14,takeiPRB14}.  A recent theoretical work has shown that such superfluid spin transport can be realized in antiferromagnetic insulators using a two-terminal setup~\cite{takeiPRB14}: by laterally attaching two strongly spin-orbit-coupled normal metals at two opposite ends of the insulator, both spin injection and detection could be achieved via electrical means using direct and inverse spin Hall effects. Transplanting this idea to the purported $\nu=0$ CAF state in graphene, superfluid transport of spin polarized along the $z$ axis could be harnessed by the CAF via a dynamic N\'eel texture that rotates about the $z$ axis within the graphene plane~\cite{takeiPRB14}. The observation of such spin superfluidity in the $\nu=0$ state should, therefore, support the above CAF scenario. 

\begin{figure}[t]
\centering
\includegraphics*[width=\linewidth]{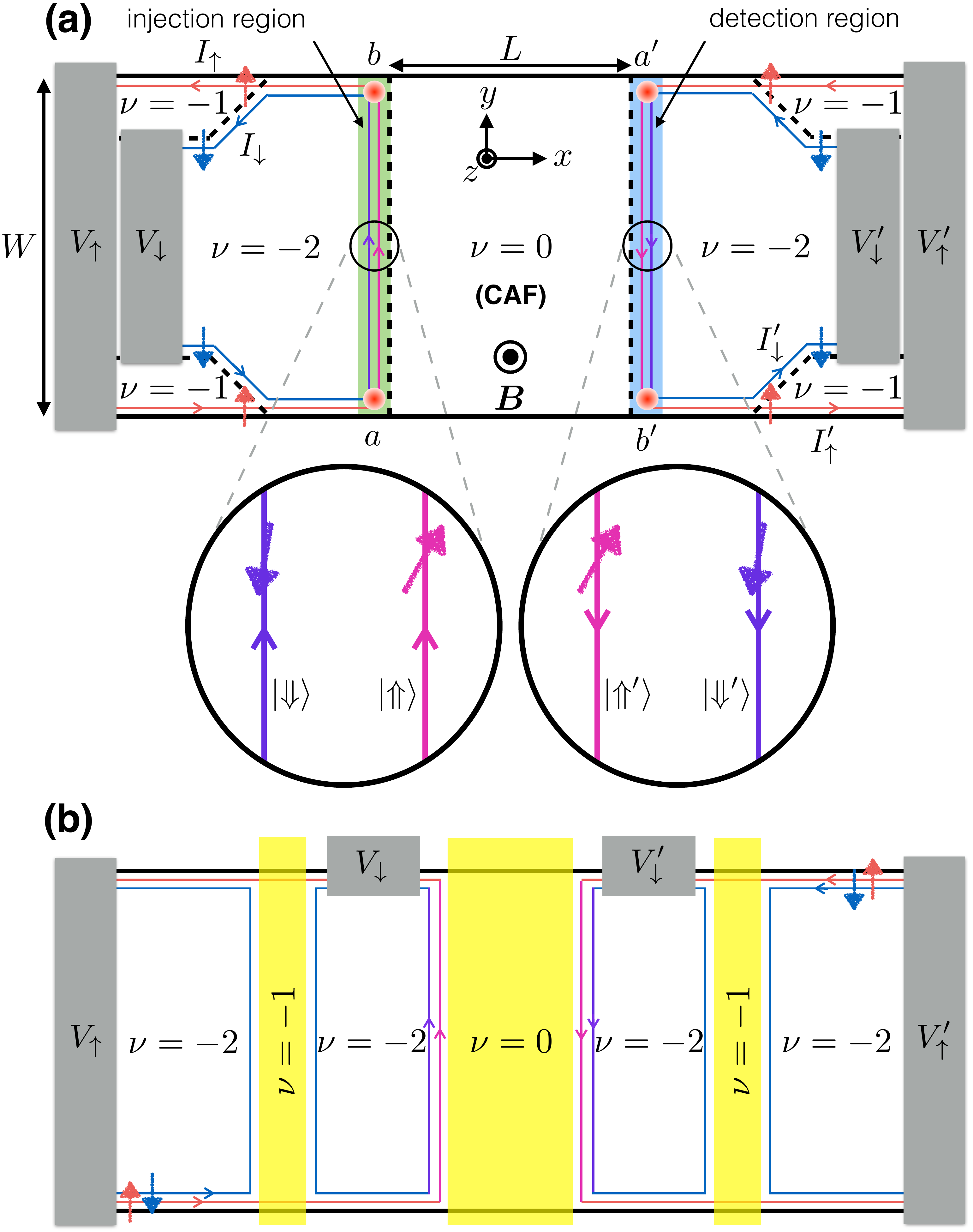}
\caption{(a) Theoretically proposed setup for realizing and detecting spin superfluid transport through the $\nu=0$ QH state of graphene. Boundaries between regions with different filling fractions are represented by dashed lines. Two independently biased spin-polarized channels on two opposite sides of the $\nu=0$ region are used to inject and detect spin current flowing through the $\nu=0$ region. The external field $\bB$ constrains the N\'eel vector to lie within the $xy$ plane. A slight misalignment of the edge spins away from the $z$ axis in the injection and detection regions may result from the neighboring CAF (see the blow-up illustration). The spin states of the $\nu=-2$ edge channels are polarized collinearly to the $z$ axis outside the injection and detection regions. (b) One possible experimental realization of the proposed setup in (a). Both spin channels biased at $V_\up$ impinge upon the gated $\nu=-1$ region, which filters one of the spin channels (the spin-down channel) from entering the inner $\nu=-2$ region. The spin-down channel within the inner region can be separately biased by $V_\down$ thereby allowing independent control of the electrochemical potentials of the two spin channels. An analogous setup is used for the detection side.}
\label{setup}
\end{figure}

{\em Superfluid spin transport}.|To realize and detect superfluid spin transport through the $\nu=0$ CAF we propose the device shown in Fig.~\ref{setup}(a), where the central CAF region is sandwiched by two $\nu=-2$ QH regions; we ignore the effects of thermal fluctuations of the spins in the CAF. Spin injection into the CAF is achieved using the two co-propagating edge channels of the left $\nu=-2$ region. Based on the theory of QH ferromagnetism~\cite{goerbigRMP11,*barlasNT12} these edge channels, away from the injection region (shaded in green), are in oppositely polarized spin states (labeled $\up,\down$) collinear with the external field (along the $z$ axis) [the injection region contains the {\em vertices}, two corners of the edge channels labeled $a$ and $b$ (represented by red circles in Fig.~\ref{setup}(a)), and the {\em line junction}, the stretch of edge channels between the vertices, adjacent to the CAF]. In the absence of spin-flip processes, the edge channels undergo very little equilibration outside the injection region~\cite{ametPRL14} so that their voltages, as they enter the region, are defined by the reservoirs from which they originate, i.e., $V_\s$. A possible experimental realization of the proposed setup is shown in Fig.~\ref{setup}(b), which also shows how the two spin channels can be separately biased. Both spin channels biased at $V_\up$ impinge upon the gated $\nu=-1$ region, which filters one of the spin channels (the spin-down channel) from entering the inner $\nu=-2$ region. The spin-down channel within the inner region can be separately biased by $V_\down$, allowing independent control of the electrochemical potentials of the two spin channels that impinge upon the injection region. An analogous setup can be used for the detection side. 

\begin{figure}[t]
\centering
\includegraphics*[width=0.85\linewidth]{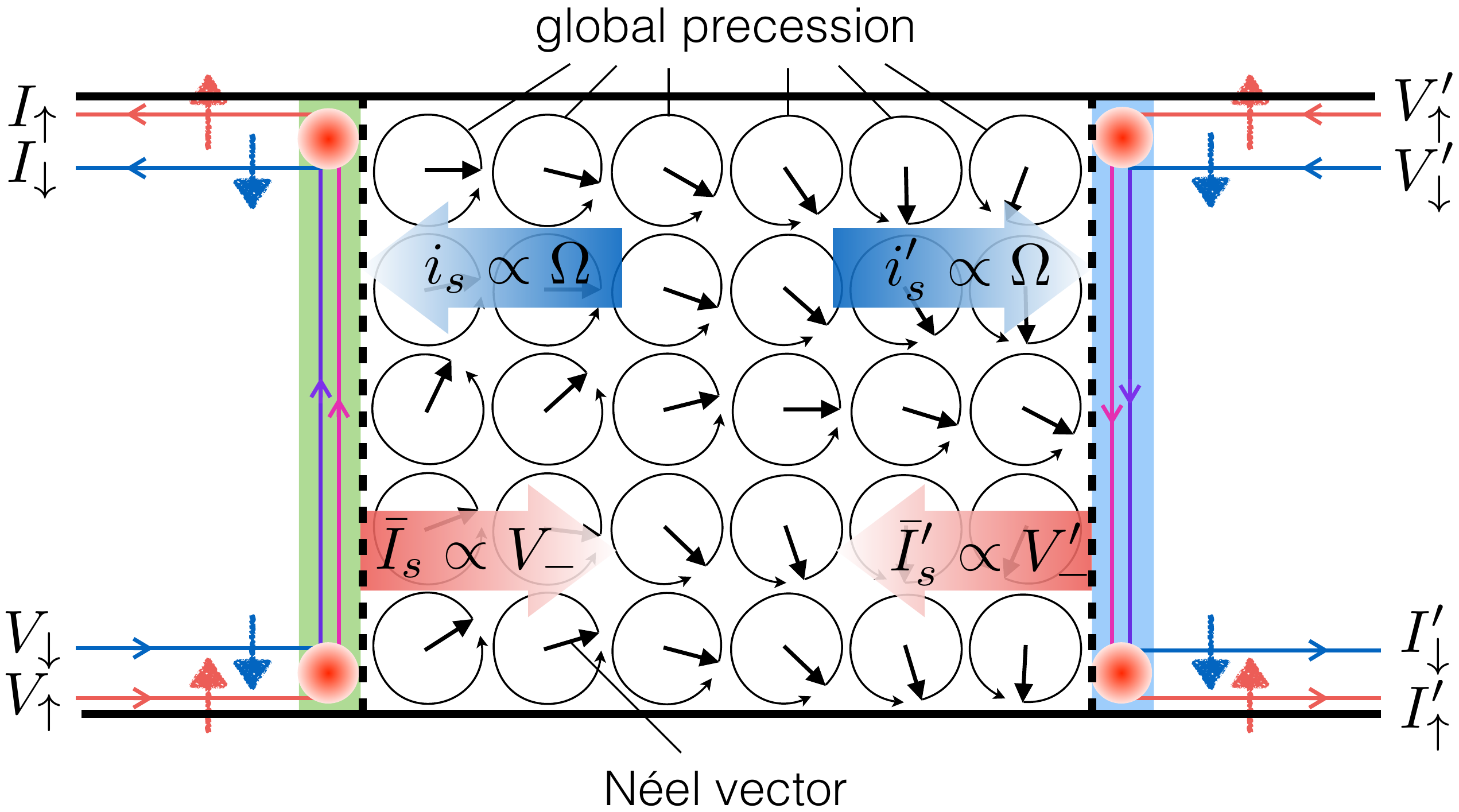}
\caption{A cartoon of the CAF in the dc superfluid state. The N\'eel vector rotates within the graphene plane about the $z$ axis with a global precession frequency $\W$, and harnesses superfluid spin transport. The phase (the azimuth) of the N\'eel vector may be spatially inhomogeneous if spin current is non-uniformly injected along the injection region. Spin current injected via the injection region has a static contribution $\bar I_s\propto V_-=V_\up-V_\down$ and a dynamic (spin-pumping) contribution $i_s\propto\W$ that pumps spin current back out into the edge. Two analogous contributions exist also on the detection side.}
\label{cartoon}
\end{figure}

When $V_\up>V_\down$, inter-channel scattering may occur inside the injection region, entailing redistributed charge currents, $I_\up$ and $I_\down$, emanating from the region and a net loss of spin angular momentum, polarized along the $z$ axis, inside the region. Neglecting any external sources of spin loss  (e.g., spin-orbit coupling, magnetic impurities, etc.) inside the injection region, the net spin lost in the edge should be fully absorbed by the CAF, leading to the injection of spin current (hereafter always defined to be the component polarized along the $z$ axis) into the CAF. This will eventually induce the CAF into a dynamic steady-state, in which the local N\'eel vector in the CAF precesses about the $z$ axis with a  global frequency $\W$ (see Fig.~\ref{cartoon})~\cite{takeiPRB14}. The dynamic N\'eel texture, in turn, pumps spin current~\cite{tserkovPRL02sp,*tserkovRMP05} out into the detection edge channels in the detection region, thus facilitating the superfluid spin transport from the injection to the detection side (the detection region, involving vertices $a'$ and $b'$, is shaded in blue in Fig.~\ref{setup}(a) and defined analogously to the injection region). Away from the detection region, the spins of the edge channels in the right $\nu=-2$ region are also oppositely polarized along the $z$ axis, so that the spin current ejected into the detection edge can be determined by measuring the difference in spin current entering and exiting the detection region.

{\em Theory and results}.|A theory for superfluid spin transport through antiferromagnetic insulators has been developed in detail in an earlier work by the authors~\cite{takeiPRB14}, and the discussion below follows directly from the work. Once the dynamic steady-state is established in the CAF, the total spin current $I_s$ entering the CAF via the injection region has two contributions: $I_s=\bar I_s+i_s$, where $\bar I_s$ is the spin current injected into a static CAF in equilibrium, and $i_s$ is the spin-pumping (dynamic) contribution describing spin current pumped back out to the edge due to the nonequilibrium N\'eel dynamics (see Fig.~\ref{cartoon})~\cite{tserkovPRL02sp,*tserkovRMP05}. Within linear-response, once the static contribution is quantified, the dynamic contribution can be determined using Onsager reciprocity, as we show below.

The static contribution to the spin current within linear-response reads $\bar I_s=(\hbar/2e)[g_Q(V_\up-V_\down)-(\bar I_\up-\bar I_\down)]$, where $g_Q\equiv e^2/h$ and $e>0$ is the magnitude of the electron charge; $\bar I_\s$ denotes the charge currents emanating from the injection region in the static limit. Due to charge conservation, and the fact that equally-biased edge channels leads to equal outgoing charge currents (i.e., $V_\up=V_\down$ implying $\bar I_\up=\bar I_\down$), the charge currents emanating from the injection region can be written generally as $\bar I_\s=g_Q[V_++\s(1-\g)V_-]/2$, where $V_{\pm}=V_{\up}\pm V_{\down}$ and $\s=\pm$ corresponds to the $\up$ and $\down$ channels, respectively. The real parameter $0\le\g\le1$ characterizes the strength of inter-channel scattering in the injection region (it is explicitly computed using a simple microscopic model at a later point in this work). The limit of no inter-channel scattering corresponds to $\g=0$, while the limit of strong scattering (full equilibration between the channels) corresponds to $\g=1$. Inserting $\bar I_\s$ into the expression for $\bar I_s$, one obtains
\beq
\label{iss}
\bar I_s=\frac{\hbar}{2e}g_Q\g V_-\, .
\eeq

The dynamical contribution $i_s$ follows from Eq.~\eqref{iss} and Onsager reciprocity. Let us first define two continuum variables in the CAF that are slowly varying on the scale of the magnetic length: $\bn(\bx)$ and $\bm(\bx)$, $\bn(\bx)$ being a unit vector pointing along the local N\'eel order and $\bm(\bx)$ being the local spin density. The global frequency $\W$ of the rotating N\'eel texture effectively acts as an additional magnetic field in the $z$ direction and introduces a uniform ferromagnetic canting of the CAF spins along the $z$ direction in addition to the existing canting due to the external field. Therefore, in the dynamic steady-state the CAF is characterized by a {\em uniform} $\bm(\bx)=m_z\ez$. Defining the total spin $M_z=m_zLW$, where $L$ and $W$ are the dimensions of the CAF region [see Fig.~\ref{setup}(a)], the dynamics of $M_z$ in the presence of the injected spin current $I_s$ is given by 
\beq
\label{dmz}
\dot M_z=I_s+\cdots\, ,
\eeq
where the ellipsis denotes terms arising from the intrinsic dynamics within the CAF. Inserting the static contribution Eq.~\eqref{iss} in for $I_s$ in Eq.~\eqref{dmz} introduces terms linear in $V_\up$ and $V_\down$, which are the forces conjugate to the charge currents $I_\up$ and $I_\down$, respectively. Onsager reciprocity then endows the static contributions $\bar I_\s$ with a dynamic contribution as
\beq
\label{isig}
I_\s=\bar I_\s-\s\frac{\hbar}{2e}g_Q\g f_{M_z}\, ,
\eeq
where $f_{M_z}\equiv -\de_{M_z} F$ is the force conjugate to $M_z$ and $F$ is the free energy of the CAF [in obtaining Eq.~\eqref{isig}, we have assumed a symmetry $\mathscr{S}$ of the device in Fig.\ref{setup}(a) under time-reversal followed by a $\p$ spatial rotation about the $x$ axis]. Noting that the force $f_{M_z}$ relates to the local N\'eel vector via $f_{M_z}=-(\bn\times\dot\bn)\cdot\ez\approx-\W$~\cite{takeiPRB14}, the total injected spin current $I_s=(\hbar/2e)[g_Q(V_\up-V_\down)-(I_\up-I_\down)]$ can be obtained using Eq.~\eqref{isig} as
\beq
\label{isfull}
I_s=\frac{\g}{4\p}(eV_--\hbar\W)\equiv\bar I_s+i_s\, .
\eeq
Based on a fully analogous consideration on the detection side, the total spin current injected into the edge from the CAF becomes $I'_s=-(\g'/4\p)(eV_-'-\hbar\W)\equiv\bar I'_s+i'_s$, where $\g'$ is the inter-channel scattering parameter, analogous to $\g$, for the detection side. Fixing the voltages of the electron reservoirs on the detection side to zero, i.e., $V'_{\up,\down}=0$, we obtain $I'_s=i'_s$.

\begin{figure}[t]
\centering
\includegraphics*[width=0.85\linewidth]{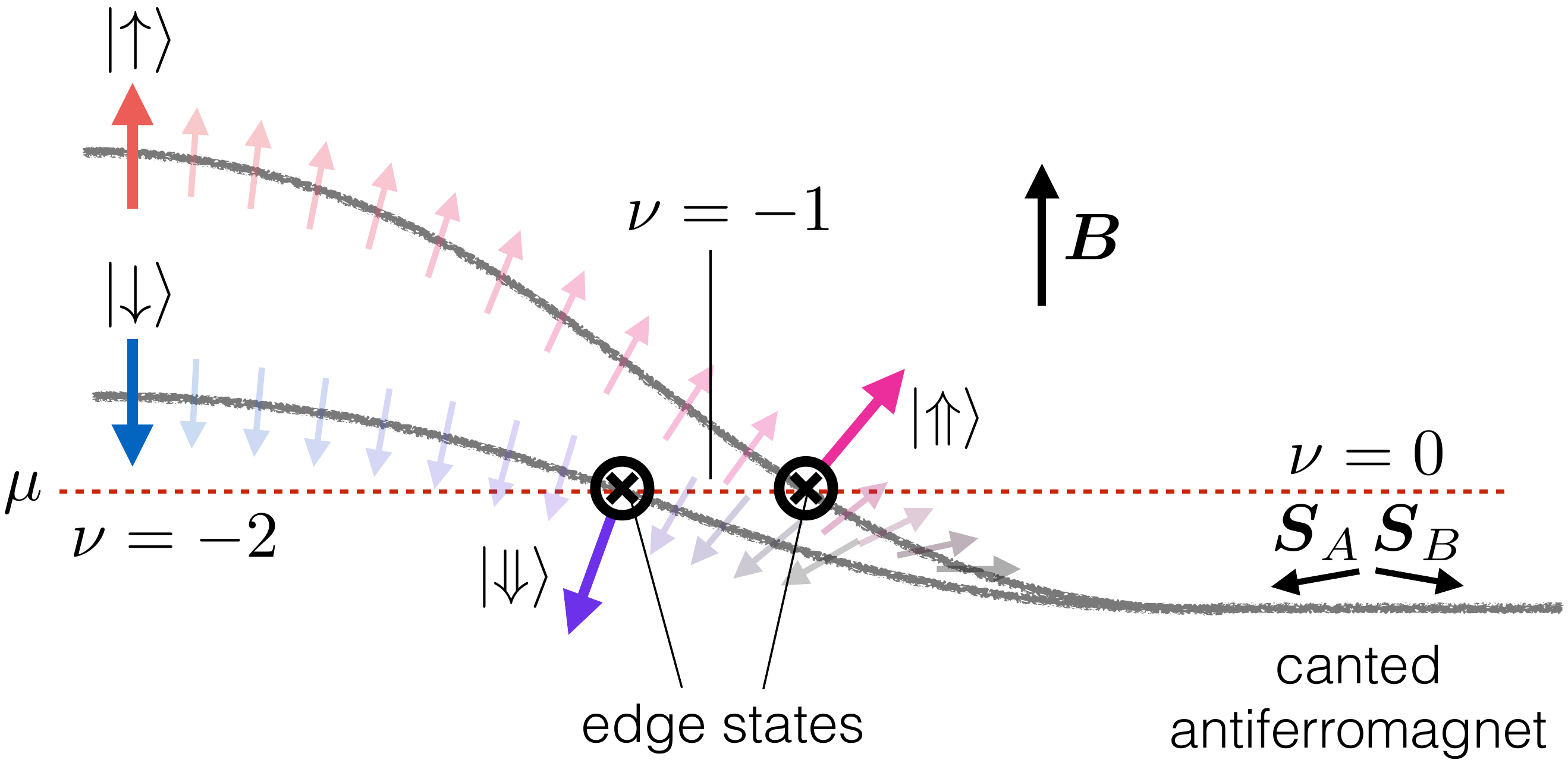}
\caption{A cartoon energy diagram at a $\nu=0$ to $\nu=-2$ transition region. In the $\nu=-2$ region, energies of the two spin states, oppositely polarized along the $z$ axis, are drawn; the Zeeman effect gives an energetic advantage to the spin-down state. In the $\nu=0$ region, the two occupied branches of the CAF spectrum are shown. There, an external field in the positive $z$ direction results in a ferromagnetic canting of spins in the negative $z$ direction inside the antiferromagnet. Spin orientations of the chiral modes are intermediate between the up and down spin eigenstates within the $\nu=-2$ region (left side) and the canted spins within the CAF (right side). The black lines are merely a rough guide for the energies of the spin states in the transition region. The above illustration does not contain two other branches of the spectrum that are a part of the zLL but not essential for the edge physics in the transition region.}
\label{edge}
\end{figure}
The dynamic N\'eel texture leads to Gilbert damping in the CAF bulk. The amount of spin current lost in the bulk reads $I_{s}-I'_{s}=\al s\W LW$~\cite{takeiPRB14}, where $\al$ is the bulk Gilbert damping parameter (whose microscopic origin is discussed at the end of the paper), and $s\equiv\hbar S/\mathscr{V}$ is the saturated spin density, with $\mathscr{V}$ denoting the area per spin of the CAF. The global frequency is then given by
\beq
\label{Omega}
\hbar\W=\frac{\g}{\g+\g'+\g_\al}eV_{-}\, ,
\eeq
where $\g_\al=4\p\al sLW/\hbar$. Then the amount of spin current generated on the detection side by the superfluid spin transport reads
\beq
\label{spincurr}
I'_{s}=\frac{1}{4\p}\frac{\g\g'}{\g+\g'+\g_\al}eV_{-}\ .
\eeq
Eqs.~\eqref{Omega} and \eqref{spincurr} constitute the main results of this work. This phenomenological result can be derived in a rotating frame, in which the spin spaces of all the edge electrons and the CAF rotate about the $z$ axis with frequency $\W$. In this frame, the voltages of the edge channels emanating from the reservoirs are shifted as $V_\s\rightarrow \tilde V_\s\equiv V_\s-\s\hbar\W/2e$ and $V'_\s\rightarrow \tilde V'_\s\equiv V'_\s-\s\hbar\W/2e$.  Since the N\'eel texture is static in the frame, the full spin current injected into the CAF on the injection side can be obtained by substituting $\tilde V_\s$ in for $V_\s$ in the expression for the {\em static} contribution, i.e., $\bar I_s$. With this substitution, one arrives directly at the result Eq.~\eqref{isfull}~\footnote{We note here that the result obtained in the rotating frame does not require the additional assumption on the symmetry $\mathscr{S}$, which was necessary in the Onsager argument. }.

\begin{figure}[t]
\centering
\includegraphics*[width=0.8\linewidth]{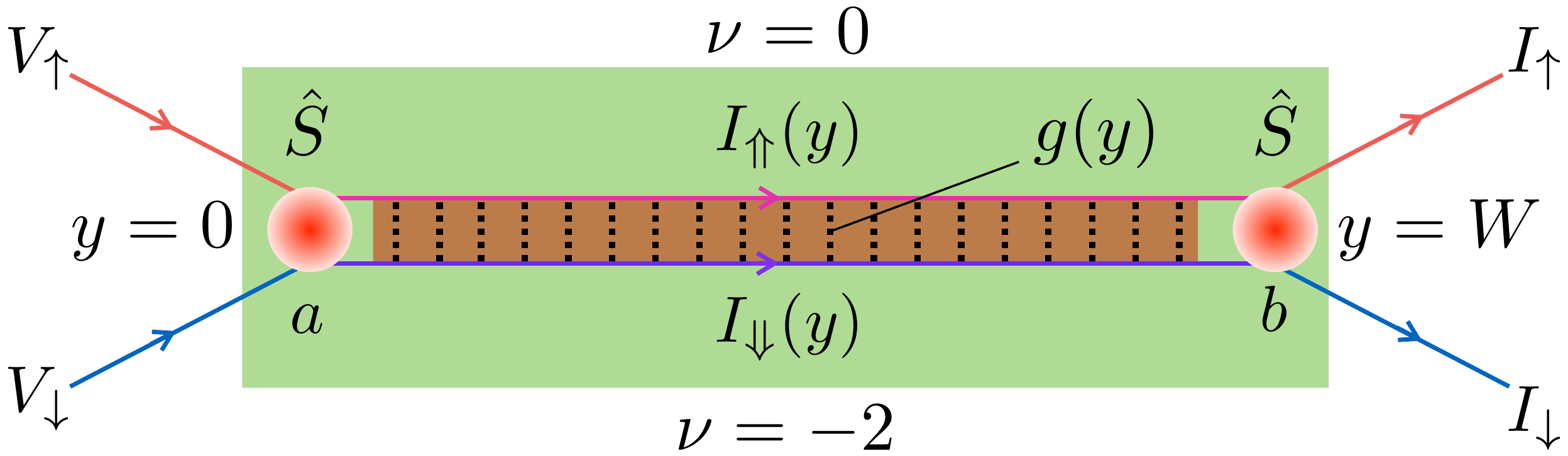}
\caption{The line junction on the injection side. Charge currents entering vertices $a$ and $b$ redistribute according to scattering probability matrix $\hat S$. Mixing of charges between the two channels in the line junction is quantified by an effective conductance per unit length $g(y)$. The green box denotes the injection region.}
\label{lj}
\end{figure}

{\em Kinetic theory for injection/detection regions}.|Here, we develop a simple microscopic model for the parameters $\g$ and $\g'$. On the injection side, $\g$ quantifies the extent to which the two edge channels equilibrate inside the injection region. Within linear-response, $\g$ can be evaluated for the (static) CAF in equilibrium. Due to the adjacent CAF order, the spin states along the line junction may deviate away from the $\pm z$ directions due to the effective field created by the CAF [see blow-up in Fig.~\ref{setup}(a) and Fig.~\ref{edge}]. The spin quantization axes there are thus expected to be canted away from the $\pm z$ axis and we label them $\Uparrow$ and $\Downarrow$. At vertices $a$ and $b$, the relative spin misalignment between the $(\up,\down)$ and $(\Uparrow,\Downarrow)$ states, together with sources of momentum non-conservation there, e.g., edge disorder and the sharp directional change of the edge, can give rise to inter-channel charge scattering. The redistribution of charges at these vertices must obey charge conservation, and can be parameterized by an energy-independent transmission probability $t\in[0,1]$ (under the assumed symmetry $\mathscr{S}$, the two vertices are characterized by an identical probability)
\beq
\label{iv}
\cvec{I_{\Uparrow}(0)}{I_{\Downarrow}(0)}=g_Q\hS\cvec{V_{\up}}{V_{\down}},\quad\cvec{I_{\up}}{I_{\down}}=\hS\cvec{I_{\Uparrow}(W)}{I_{\Downarrow}(W)},
\eeq
where $I_{\bbsigma}(y)$ (with $\bbsigma=\Uparrow,\Downarrow$) is the local charge current flowing along the line junction in edge channel $\bbsigma$, $\hS=t\hat\s_0+(1-t)\hat\s_x$ is the scattering probability matrix at the vertices, and $\hat\s_0$ and $\hat\s_x$ are the $2\times2$ identity matrix and the $x$ component of the Pauli matrices, respectively. 

For inter-channel scattering inside the line junction, one requires: (i) spatial proximity of the two channels, such that there is sufficient overlap of their orbital wave functions; (ii) elastic impurities, providing the momentum non-conserving mechanism necessary to overcome the mismatch in Fermi momenta of the two channels; and (iii) a spin-flip mechanism, assumed here to be provided by the neighboring CAF. All three factors go into defining the tunneling conductance $g(y)$ per unit length between the edge channels. In terms of $g(y)$, the change in current on channel $\bbsigma$ is given by $\de I_{\Uparrow,\Downarrow}(y)=\mp g(y)[V_{\Uparrow}(y)-V_{\Downarrow}(y)]\de y$, where $V_{\bbsigma}$ is the local voltage on edge channel $\bbsigma$ [we assume that the edges are always locally equilibrated at all points $y$ such that the voltage at each point is related to the local current through $V_{\bbsigma}(y)=I_{\bbsigma}(y)/g_Q$]. Then, the currents inside the line junction satisfy 
\beq
\frac{\pd I_{\Uparrow}}{\pd y}=-\frac{\pd I_{\Downarrow}}{\pd y}=-\frac{g(y)}{g_Q}[I_{\Uparrow}(y)-I_{\Downarrow}(y)].
\eeq
Assuming a position-independent tunneling conductance $g$ and defining the edge equilibration length $\ell\equiv g_Q/2g$, the currents entering vertex $b$ is then given by
\beq
\label{ilj1}
\cvec{I_{\Uparrow}(W)}{I_{\Downarrow}(W)}=\frac{1}{2}\mat{1+e^{-W/\ell}}{1-e^{-W/\ell}}{1-e^{-W/\ell}}{1+e^{-W/\ell}}\cvec{I_{\Uparrow}(0)}{I_{\Downarrow}(0)}\ .
\eeq
Combining Eqs.~\eqref{iv} and \eqref{ilj1}, the parameter $\g$ (on the injection side) reads
\beq
\label{g}
\g=1-(1-2t)^2e^{-W/\ell}.
\eeq
A fully analogous consideration on the detection side leads to $\g'=1-(1-2t')^2e^{-W/\ell'}$, where $t'$ is the transmission probability at vertices $a'$ and $b'$, and $\ell'$ is the edge equilibration length associated with the line junction on the detection side.

\begin{figure}[t]
\centering
\includegraphics*[width=0.9\linewidth]{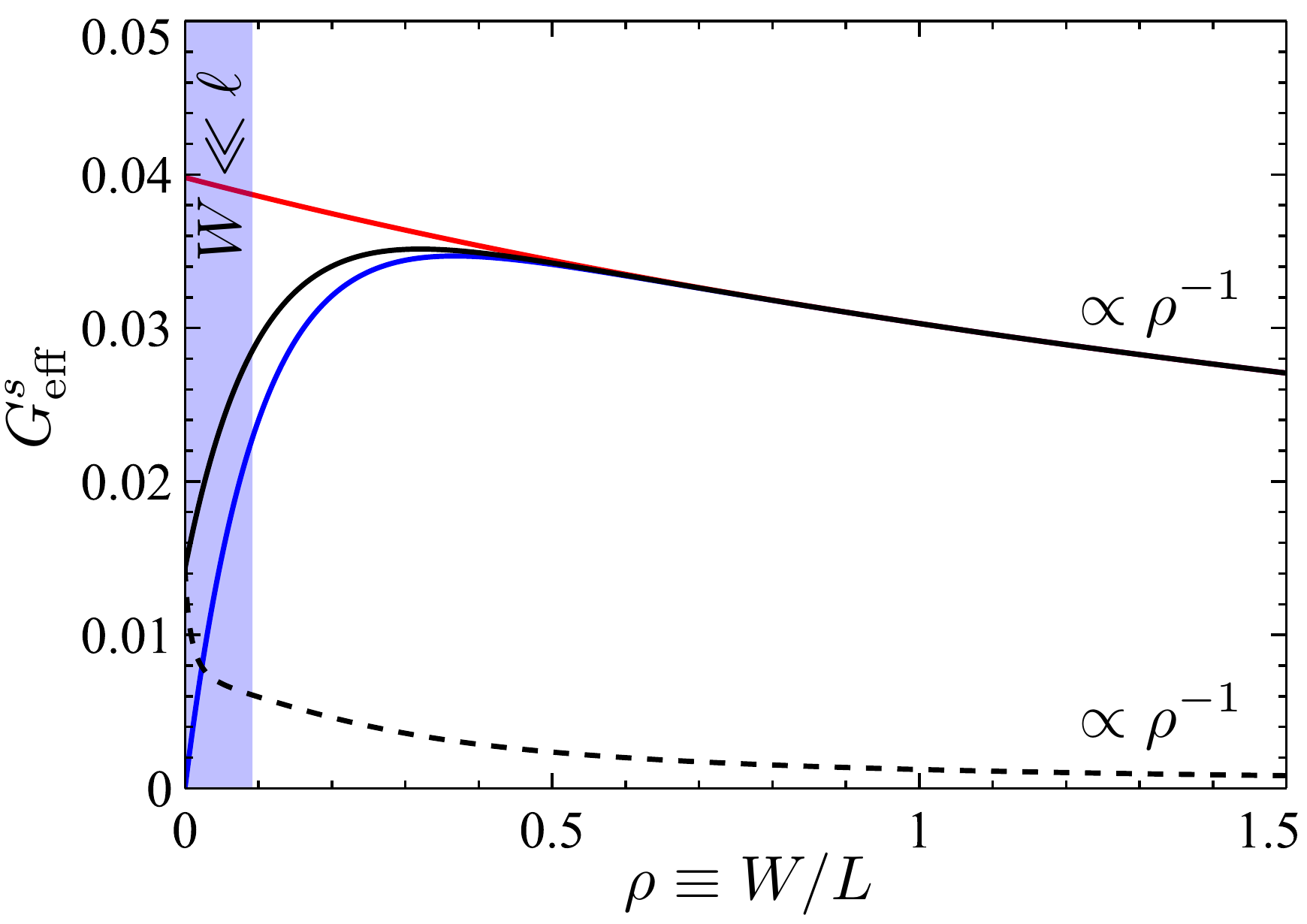}
\caption{Effective spin conductance $G^s_{\rm eff}\equiv I'_s/eV_-$ as a function of the aspect ratio $\rho\equiv W/L$. We fix the edge equilibration length $\ell$ and the system length $L$ so that $l\equiv L/\ell=10$. The red and blue curves are, respectively, for full ($t=1/2$) and no ($t=1$) inter-channel mixing at the vertices, and effective Gilbert damping $\al_{\rm eff}=0.05$ is used. The solid and dashed black lines, respectively, correspond to the weak ($\al_{\rm eff}=0.05$) and strong ($\al_{\rm eff}=5$) Gilbert damping with a transmission probability at the vertices of $t=0.9$. }
\label{sc}
\end{figure}

{\em Discussion}.|For clarity, the results are now discussed for the symmetric case, in which the injection and detection sides are characterized by an identical transmission probability and edge equilibration length, i.e., $t=t'$ and $\ell=\ell'$. Recall that $\ell$ describes the length scale over which the two relatively-biased edge channels in the line junctions chemically equilibrate via inter-channel scattering; we fix $\ell$ and the length of the CAF region $L$ to the ratio $l\equiv L/\ell=10$. The effective spin conductance through the CAF, $G^s_{\rm eff}\equiv I'_{s}/eV_{-}$ [see Eq.~\eqref{spincurr}], can then be expressed in terms of two variables: the aspect ratio $\rho\equiv W/L$ and the effective Gilbert damping $\al_{\rm eff}\equiv \al s L^2/\hbar$ in the bulk CAF. 

The effective spin conductance is plotted as a function of the aspect ratio $\rho$ for different $t$ and $\al_{\rm eff}$. Full mixing of the edge channels at the vertices, i.e., $t=1/2$, entails spin current injection only at vertex $a$. In this case, increasing $\rho$ only increases the effects of Gilbert damping, since the latter is a bulk effect of the CAF, and the spin conductance monotonically decreases essentially as $G^s_{\rm eff}\propto\rho^{-1}$ (see the red line). We call this ``damping-dominated" behavior. If no scattering occurs at the vertices, i.e., $t=1$, spin current can only be injected within the line junction. For widths much smaller than the equilibration length, i.e., $\rho l\ll1$, increasing the width gives an enhancement in the injected spin current that overcomes losses due to Gilbert damping, and a linear increase $G^s_{\rm eff}\propto\rho$ (see the blue line) is obtained. However, as the width increases beyond the equilibration length, spin injection no longer increases while Gilbert losses continue to increase. This leads to the eventual decay $G^s_{\rm eff}\propto\rho^{-1}$ for large $\rho$. We call this ``weak-damping" behavior. 

For partial inter-channel mixing at the vertices, $0<t<1$, $G^s_{\rm eff}$ has a qualitatively different dependence on $\rho$ for different Gilbert damping strengths. Let us consider device widths much less than the edge equilibration length, $W\ll\ell$ (the blue shaded region in Fig.~\ref{sc}). Here, the Gilbert damping effects are small as long as the effective spin conductance in the lossless regime, characterized by $\g$, is much larger than the effective spin conductance $\g_\al$ associated with Gilbert damping. For $\g\gg\g_\al$, $G^s_{\rm eff}$ exhibits the weak-damping behavior where an enhancement in the spin conductance is observed for $W\ll\ell$ (see the solid black curve). Damping-dominated behavior is restored for stronger Gilbert damping, i.e., $\g\ll\g_\al$, where the effective spin conductance monotonically decreases with the aspect ratio (see the dashed black curve). 

The Gilbert damping parameter $\al$, as introduced in Eq.~\eqref{Omega}, quantifies macroscopic relaxation of spin angular momentum polarized along the $z$ axis, arising only in the presence of both spin-orbit interactions and microscopic degrees of freedom for energy dissipation. For a perfect graphene membrane, the spin-orbit coupling is known to be weak, while energy dissipation may be provided by phonons and magnons. Gilbert damping should vanish in this case as one approaches zero temperature, as phonons and magnons freeze out. Magnetic or heavy-element impurities and/or enhanced spin-orbit interactions due to substrate can increase the spin relaxation rate. It is known, for instance, that spin-orbit interactions in graphene can be enhanced, e.g., by hydrogenation~\cite{balakrishnanNATP13} or interfacing it with a heavy-element$-$based semiconductor~\cite{avsarNATC14}. Additional dissipation channels, furthermore, can stem from substrate phonons or an ensemble of two-level systems (rooted in, e.g., crystalline defects either in graphene or the substrate). The latter could lead to energy relaxation (and thus finite $\al$) even as $T\rightarrow0$~\cite{gaoPHD08,*tabuchiPRL14}. In a flat graphene membrane, U(1) symmetry-breaking magnetocrystalline anisotropies, which, in principle, could pin the orientation of the in-plane staggered magnetization and thereby quench the superfluid state~\cite{soninAP10,takeiPRL14,takeiPRB14}, are expected to be small due to graphene's weak spin-orbit coupling and high space-group symmetry of the hexagonal lattice.

\begin{acknowledgments}
{\em Acknowledgments}.|We would like to thank Dmitry Abanin and Jelena Klinovaja for helpful discussions. This work was supported by FAME (an SRC STARnet center sponsored by MARCO and DARPA). B.~I.~H. and A.~Y. were supported in part by the STC Center for Integrated Quantum Materials under NSF Grant No. DMR-1231319. 
\end{acknowledgments}

\end{document}